\documentstyle[wstwocl,epsfig]{article}
\begin{document}

\bibliographystyle{unsrt}    

\newcommand{\st}{\scriptstyle}
\newcommand{\sst}{\scriptscriptstyle}
\newcommand{\mco}{\multicolumn}
\newcommand{\epp}{\epsilon^{\prime}}
\newcommand{\vep}{\varepsilon}
\newcommand{\ra}{\rightarrow}
\newcommand{\ppg}{\pi^+\pi^-\gamma}
\newcommand{\vp}{{\bf p}}
\newcommand{\ko}{K^0}
\newcommand{\kb}{\bar{K^0}}
\newcommand{\al}{\alpha}
\newcommand{\ab}{\bar{\alpha}}
\def\be{\begin{equation}}
\def\ee{\end{equation}}
\def\bea{\begin{eqnarray}}
\def\eea{\end{eqnarray}}
\def\CPbar{\hbox{{\rm CP}\hskip-1.80em{/}}}

\def\ap#1#2#3   {{\em Ann. Phys. (NY)} {\bf#1} (#2) #3.}
\def\apj#1#2#3  {{\em Astrophys. J.} {\bf#1} (#2) #3.}
\def\apjl#1#2#3 {{\em Astrophys. J. Lett.} {\bf#1} (#2) #3.}
\def\app#1#2#3  {{\em Acta. Phys. Pol.} {\bf#1} (#2) #3.}
\def\ar#1#2#3   {{\em Ann. Rev. Nucl. Part. Sci.} {\bf#1} (#2) #3.}
\def\cpc#1#2#3  {{\em Computer Phys. Comm.} {\bf#1} (#2) #3.}
\def\err#1#2#3  {{\it Erratum} {\bf#1} (#2) #3.}
\def\ib#1#2#3   {{\it ibid.} {\bf#1} (#2) #3.}
\def\jmp#1#2#3  {{\em J. Math. Phys.} {\bf#1} (#2) #3.}
\def\ijmp#1#2#3 {{\em Int. J. Mod. Phys.} {\bf#1} (#2) #3.}
\def\jetp#1#2#3 {{\em JETP Lett.} {\bf#1} (#2) #3.}
\def\jpg#1#2#3  {{\em J. Phys. G.} {\bf#1} (#2) #3.}
\def\mpl#1#2#3  {{\em Mod. Phys. Lett.} {\bf#1} (#2) #3.}
\def\nat#1#2#3  {{\em Nature (London)} {\bf#1} (#2) #3.}
\def\nc#1#2#3   {{\em Nuovo Cim.} {\bf#1} (#2) #3.}
\def\nim#1#2#3  {{\em Nucl. Instr. Meth.} {\bf#1} (#2) #3.}
\def\np#1#2#3   {{\em Nucl. Phys.} {\bf#1} (#2) #3.}
\def\pcps#1#2#3 {{\em Proc. Cam. Phil. Soc.} {\bf#1} (#2) #3.}
\def\pl#1#2#3   {{\em Phys. Lett.} {\bf#1} (#2) #3.}
\def\prep#1#2#3 {{\em Phys. Rep.} {\bf#1} (#2) #3.}
\def\prev#1#2#3 {{\em Phys. Rev.} {\bf#1} (#2) #3.}
\def\prl#1#2#3  {{\em Phys. Rev. Lett.} {\bf#1} (#2) #3.}
\def\prs#1#2#3  {{\em Proc. Roy. Soc.} {\bf#1} (#2) #3.}
\def\ptp#1#2#3  {{\em Prog. Th. Phys.} {\bf#1} (#2) #3.}
\def\ps#1#2#3   {{\em Physica Scripta} {\bf#1} (#2) #3.}
\def\rmp#1#2#3  {{\em Rev. Mod. Phys.} {\bf#1} (#2) #3.}
\def\rpp#1#2#3  {{\em Rep. Prog. Phys.} {\bf#1} (#2) #3.}
\def\sjnp#1#2#3 {{\em Sov. J. Nucl. Phys.} {\bf#1} (#2) #3.}
\def\spj#1#2#3  {{\em Sov. Phys. JEPT} {\bf#1} (#2) #3.}
\def\spu#1#2#3  {{\em Sov. Phys.-Usp.} {\bf#1} (#2) #3.}
\def\zp#1#2#3   {{\em Zeit. Phys.} {\bf#1} (#2) #3.}

\setcounter{secnumdepth}{2} 


\title{INFLUENCE OF ADDITIONAL FERMIONS AND GAUGE BOSONS ON
$Re(\epsilon'/\epsilon)$}

\firstauthors{Paul H. Frampton}

\firstaddress{Institute of Field Physics, Department of Physics and Astronomy,
University of North Carolina, Chapel Hill, NC 27599-3255, USA}

\twocolumn[\maketitle\abstracts{
The CP violation parameter $Re(\epsilon'/\epsilon)$
has previously been calculated in the standard model and the result depends on
the top quark mass in an exciting way. We consider how the result
is modified in two specific extensions of the standard model.
The first adds only fermions, an extra vector-like quark doublet, and tends
to reduce $Re(\epsilon'/\epsilon)$. The second contains three extended families
with the third family treated differently and additional gauge bosons -
the 331 model - and can accommodate larger values of $Re(\epsilon'/\epsilon)$
for a given $m_t$. When the experimental values of $m_t$ and $Re(\epsilon'/
\epsilon)$ are known more accurately, it will allow discrimination between
theories. If the lower experimental number $Re(\epsilon'/\epsilon)
= (0.74 \pm 0.60) \times 10^{-3}$ (Fermilab, E731) is confirmed
and $m_t$ is near the value 175GeV the standard model is acceptable. But,
if the higher value $Re(\epsilon'/\epsilon) = (2.3 \pm 0.65) \times 10^{-3}$
(CERN, NA31) persists, a dynamics like the 331 model where the third family
is treated differently from the first two and
there exist additional gauge bosons seems to be favored.}]
\section{Introduction}
Symmetry under the combined operation of C (charge conjugation)
and P (parity) holds only approximately for weak interactions. Violation of
CP was a surprising discovery made in 1964 in the neutral kaon system.
Since then the effect still has not been seen in any other system
and is parametrised by the quantities $\epsilon$ and
$Re(\epsilon'/\epsilon)$. The first parameter is known experimentally quite
accurately and given as:

\begin{equation}
\epsilon = (2.26 \pm 0.09) \times 10^{-3} exp(i\phi(\epsilon))
\end{equation}

\noindent
where
\begin{equation}
\phi(\epsilon) = 43.68 \pm 0.14degrees.
\end{equation}

The experimental measurements of the second parameter $Re(\epsilon'/\epsilon)$
are less settled as stated in the abstract above.

In 1973 the KM mechanism was introduced. Following that, a long series of
estimates have been made of the quantity $Re(\epsilon'/\epsilon)$,
until the late 1980's working with the assumption $m_t < M_W$ and
subsequently with the correct inequality. Given $m_t > M_W$, it
was found that  the electroweak penguin diagrams are large and
tend to cancel the gluon penguin diagrams. The cancellation becomes complete
for $m_t$ about 220GeV.

Thus the high $m_t$ leads in the standard model to a small value of
$Re(\epsilon'/\epsilon)$ close to the zero value predicted by the
superweak theory.

The first extension of the standard model we consider
is the rather ordinary one where we simply add a fourth family with
vector-like quarks. This does not effect agreement with precision
electroweak measurements (S,T parameters). We calculate for this case
including QCD and electroweak corrections.

The other extension we consider is the 331 model which contains new fermions
- the exotic quarks D, S and T - and new gauge bosons - the
dileptons $(Y^{--},Y^{-})$,$(Y^{++},Y^{+})$ and a $Z'$. In this case,
the calculation of $Re(\epsilon'/\epsilon)$ is far richer in the
number of Feynman diagrams. We find a generalized GIM mechanism
for the dileptons. The result is calculated as a function of
the dimensionless ratios $x_D = M_D^2/M_Y^2$ ( we always assume $M_D = M_S$)
and $x_T = (M_T^2/M_Y^2)$.  The results depend sensitively on these ratios and
on $m_t$.

The calculation involves several steps and theoretical uncertainties. While the
short-range
physics is well controlled by perturbation theory, the hadronic matrix elements
involve confinement
physics of QCD and are naturally less accurately estimated. Nevertheless, such
uncertainties
should not effect the general conclusions for specific
models.

\section{Description of the calculation of $Re(\epsilon'/\epsilon)$}

Here we outline the basic steps needed in the calculation of
$Re(\epsilon'/\epsilon)$
in a general model which {\it may} extend the standard model. For the
calculation of the effective Hamiltonian
at low energy (1GeV) we integrate out the heavy quarks by using RG equations.
One calculates the Feynman diagrams at $M_W$
scale to find the initial values of the Wilson coefficients which are produced
by the strong
and electroweak corrections to the original ($ \Delta S = 1$) operators.
Thus, after integrating the RGEs between quark thresholds and dropping
heavy quarks at each stage, one gets an effective hamiltonian
at $\mu = 1GeV$ scale as

\begin{equation}
H_{eff}(\mu) = -G_F/\sqrt2 \sum^8_{i=1,i\neq4} [C^c_i(\mu) +
\lambda_t(C^t_i(\mu)-C^c_i(\mu)]Q_i
\end{equation}
where $\lambda_t = V_{ts}^*V_{td}$, and the superscripts refer to the quark
flavors.
Substituting the effective hamiltonian into the definition of
$Re(\epsilon'/\epsilon)$
one obtains an expression in terms of the Wilson coefficients $C_i$ and the
hadronic matrix elements of operators $Q_i$.

The calculation can be conveniently divided into four steps:

(1) Finding the Wilson coefficients at the $M_W$ scale.

(2) Bringing down the energy scale to $\mu < m_c$ from $\mu = M_W$ by using the
renormalization group
equations.

(3) Estimating the hadronic matrix elements.

(4) Calculating the phase parameter in the CKM matrix, $\delta$.

\subsection{Initial value of Wilson coefficients}

The original four-quark operator corresponding to $ \Delta S = 1$ generates
a new set of four-quark operators through the box diagrams, B(x); the Z-penguin
diagrams,
C(x); the photon penguin diagrams, D(x); and the gluon penguins, E(x). At
one-loop
order, one finds a complete set of four-quark operators generated by these
diagrams and
by additional diagrams, if any.

The expressions for the functions B(x), etc. in the standard model as well as
the resulting values for the Wilson coefficients $C^{t,c}(M_W)$ exist in the
literature.
We calculate them for the extensions of the standard model.

\subsection{Wilson coefficients at $\mu = 1GeV$}

The evolution of the Wilson coefficients from the energy scale
$\mu = M_W$ to energy scale $\mu = m_c$ are governed by the RGEs which
are soluble in terms of an evolution operator which requires
the evaluations of all the relevant anomalous dimensions.

\subsection{Hadronic matrix elements.}

The effective hamiltonian after the decoupling of heavy quark states contains
seven operators
whose matrix elements need to be evaluated. This can be done using any one of
several methods
such as QCD sum rules, lattice gauge theories, factorization, or the 1/N
expansion.

We follow the 1/N expansion approach and the matrix elements using this are
given as:

\begin{equation}
<Q_1>_0 = (1/3)(2/N - 1)XB_1, etc.
\end{equation}
\begin{equation}
<Q_1>_2 = <Q_2>_2 = \sqrt2/3(1 + 1/N)XB_{27},  etc.
\end{equation}
\noindent
where $B_{1,27......}$ are bag parameters.

\subsection{Calculation of $\delta$ in CKM matrix}

The phase parameter in the CKM matrix is obtained using the experimental value
of $\epsilon$. By definition

\begin{equation}
\epsilon = e^{(i\pi /4)} (ImM_{12} + 2\xi ReM_{12})/(\sqrt2 \Delta M)
\end{equation}
\noindent
where

\begin{equation}
\xi = ImA_0/ReA_0
\end{equation}
\noindent
$M_{12}$ is the off-diagonal matrix element in the neutral K-meson mass matrix
and
$\Delta M$ is the $K_1-K_2$ mass difference.
The theoretical expression for $M_{12}$ is evaluated from the usual box diagram
in
terms of the "bag" parameter $B_K$.

\section{Model with Additional Vector-like Quarks}

We consider the extension of the standard model which contains a "fourth
family" of quarks. Thus we simply add one $SU(2)$ doublet of vector-like
quarks:

$$\left( \begin{array}{c} U^{\alpha} \\ D^{\alpha} \end{array} \right)_L
\left( \begin{array}{c} U^{\alpha} \\ D^{\alpha} \end{array} \right)_R  .$$

Only the left-handed doublet participates and makes contributions to all the
Feynman diagrams
of the standard model {\it i.e.} box, gluon-penguin, Z-penguin and
photon-penguin. The new particles
do not lead to diagrams of new topologies.

This is similar to a fourth generation except that the anomalies
have been cancelled directly by the quarks and we do not need to consider
an additional charged lepton or a massive neutrino.

The functions $B(x_t,x_Q)..........$ are, of course, functions of the mass
$M_Q$ of the extra quarks.

The result may be summarised by saying that the value of
$Re(\epsilon'/\epsilon)$
is always made {\it more negative}. Thus, if $Re(\epsilon'/\epsilon)$ turns out
to be zero or even negative, the inclusion of extra fermions is favored. On the
other hand, if
the answer is as large as the central CERN result, then additional fermions
alone will
not accommodate the empirical value.

\section{Model with Additional Gauge Bosons}

In the present section, we shall study a richer and more interesting extension
which contains both additional
fermions and gauge bosons.

In particular, we use the 331 model which is motivated by accommodation of
three families. The model has gauge group $SU(3)_c \times SU(3)_L \times
U(1)_X$ where
$SU(3)_L \times U(1)_X$ contains the standard electroweak gauge group. The
larger gauge group
is broken at a scale U by the vacuum expectation of a Higgs triplet.

The fermions transform as follows. The quarks of the first and second families
are in left-handed triplets and
right-handed singlets of $SU(3)_L$,

$$\left( \begin{array}{c} u^{\alpha} \\ d^{\alpha} \\ D^{\alpha} \end{array}
\right)_L     \bar{u}_{\alpha L}
  \bar{d}_{\alpha L}  \bar{D}_{\alpha L}   $$
\noindent
and similarly for c, s, S. The third family of quarks is in a left-handed
anti-triplet $3^*_L = (b, t, T)$ and right-handed singlets.

The leptons for all three families are in anti-triplets:

$$\left( \begin{array}{c} e^{-} \\ \nu_e \\ e^{+} \end{array} \right)_L
\left( \begin{array}{c} \mu^{-} \\ \nu_{\mu} \\ \mu^{+} \end{array} \right)_L
\left( \begin{array}{c} \tau^{-} \\ \nu_{\tau} \\ \tau^{+} \end{array}
\right)_L$$

\noindent
All anomalies cancel with three families, although not for each family
separately.

The additional five gauge bosons are a $Z'$ and two doublets of dileptons
$(Y^{--}, Y^-)$, $(Y^{++}, Y^+)$ with lepton numbers $L = +2, -2$ respectively.
The breaking
by $<\Phi_3> = U$ gives masses to $Z'$ and Y by the Higgs mechanism. The quarks
$Q = D, S, T$ also acquire mass from $<\Phi>$ through their Yukawa couplings.
We shall take $M_D = M_S$ for simplicity but allow $M_T$ to vary independently.
The dilepton mass $M_Y$ must be above 300GeV but can be up to 1100GeV. We
denote $x_D
= M_D^2/M_Y^2$ and $x_T = M_T^2/M_Y^2$ and allow them to vary in the range $0.1
\leq x_{D,T}
\leq 10.$

The new particles introduce additional Feynman diagrams beyond the usual
standard model
for $K \rightarrow \pi\pi$ decay at one-loop level. The ones of the same
topology
modify the functions B, C, D, E which now depend on $x_D$ and $x_T$. A new
topology
enters through the mixings of $Z'$ with $\gamma, Z$ but this is used only to
cancel
certain divergences in the electroweak penguin diagrams, the finite part being
very
small due to the large $Z'$ mass.

After evaluation of the modified B, C, D, E functions, it is interesting to
note the following
points:

(1) For box and gluon-penguin diagrams, the infinities are cancelled by a
generalized
GIM mechanism and the functional dependence of these diagrams on the respective
$x_Q$ turns
out to be the same as results in the standard model.

(2) The coupling between the exotic quarks and the $Z$ boson is vector-like. As
a
consequence of this and of gauge invariance, the finite part of the Z-penguin
vanishes at order
$(q^2/M_Z^2)$, where $q_{\mu}$ is the four-momentum transfer.

(3) Though the generalized GIM mechanism does not remove completely the
divergences of the photon-penguin,
they do nevertheless cancel as a result of gauge invariance after including the
new-topology
diagrams mentioned above. Another interesting point in the photon-penguin is
that all three
exotic quarks contribute with the same sign: the charge in the closed loop
flows in the opposite
sense for the T relative to D and S and negates the sign change normally
arising
from the unitarity of the CKM matrix. This is a result of the asymmetric
treatment of the
third family, and is crucial in obtaining a more positive result for
$Re(\epsilon'/\epsilon)$.

The result for $Re(\epsilon'/\epsilon)$ is that the minimum value is
essentially identical,
within one per cent, to the standard model. The maximum value can accommodate
$2.3 \times 10^{-3}$
if we allow acceptable choices of $\Lambda_{\overline{MS}}$ and $m_s$.

\section{Discussion and Summary}

The values of the direct CP violation parameter is generally non-zero if there
is direct
$\Delta S = 1$ CP violation as in the standard model; it is very small $\ll
10^{-4}$
only from delicate cancellations.

In the standard model, the prediction for $Re(\epsilon'/\epsilon)$ was thought
to be
about $+3 \times 10^{-3}$ when $m_t \ll M_W$ was assumed, and the Z-penguins
were neglected. With the heavy top
quark, and inclusion of Z-penguins, the prediction is now in the range from
$0.8 \times 10^{-3}$
down to zero if we take conservative values for all the other parameters.

We have considered how
$Re(\epsilon'/\epsilon)$ can be effected by additional fermions and gauge
bosons.

What we find is that the simplest additional fermions tend to decrease the
prediction
compared to the standard model. On the other hand, the extension which treats
the third
family differently can increase the prediction to a value $2 \times 10^{-3}$.

The future should bring us accurate experimental values, both for
$Re(\epsilon'/\epsilon)$
as well as for the CKM parameters, $\Lambda_{\overline{MS}}$, $m_s$ and $B_K$.
As further information on $Re(\epsilon'/\epsilon)$ emerges resolving the
discrepancy
between the current Fermilab $(0.74 \times 10^{-3})$ and CERN $(2.3 \times
10^{-3})$
central values, it will indicate not only whether new physics is involved but
what form is
required.

If the smaller Fermilab value for $Re(\epsilon'/\epsilon)$ is verified, the
standard
model can accommodate it. If a vanishingly small value emerges, superweak
theory will be
favored. If a negative value appears, extra fermions are indicated.

If the larger CERN central value is confirmed then it hints that there is
something beyond the standard model, and of the scenarios we have considered
the most favored is that with additional gauge bosons and an asymmetrical
treatment
of the third generation.

The full calculation is published in:
J. Agrawal and P. H. Frampton, Nucl. Phys. {\bf B419}, 254-278 (1994)
The U.S. Department of Energy is thanked for support.
An extensive bibliography, for which inadequate space remains, is provided
in the Nucl. Phys. {\bf B} article.

\end{document}